\documentclass[%
 reprint,
superscriptaddress,
 amsmath,amssymb,
 aps,
pra,
]{revtex4-2}

\usepackage{graphicx}
\usepackage{dcolumn}
\usepackage{bm}

\usepackage{amsmath}
\usepackage{lipsum}
\usepackage{upgreek}
\DeclareMathOperator{\sinc}{sinc}

\usepackage{braket}

\usepackage{graphicx}
\usepackage{dcolumn}
\usepackage{bm}
\usepackage[colorlinks,linkcolor=blue,citecolor = blue,urlcolor=blue]{hyperref}
\usepackage{booktabs}
\usepackage{float}
\usepackage{csquotes}
\usepackage{abraces}

\usepackage{commath}
\usepackage{ulem}
\usepackage{braket}
\usepackage{color}
\usepackage{cleveref}
\usepackage{bbold}
\usepackage{booktabs}
\usepackage{gensymb}
\usepackage[abs]{overpic}
\usepackage{mathtools}
\usepackage{graphicx}

\begin{document}

\preprint{APS/123-QED}


\title{Mode-selective nonlinear interference for high-brightness and high-purity fiber-coupled SPDC sources}


\author{Carlos Sevilla-Gutiérrez}\email{carlos.sevilla@iof.fraunhofer.de}
\affiliation{Fraunhofer Institute for Applied Optics and Precision Engineering, 
07745 Jena, Germany}
\affiliation{Friedrich Schiller University Jena, Institute of Applied Physics, Abbe Center of Photonics, 
07745 Jena, Germany}
\author{Purujit Singh Chauhan}
\affiliation{Fraunhofer Institute for Applied Optics and Precision Engineering, 
07745 Jena, Germany}
\affiliation{Friedrich Schiller University Jena, Institute of Applied Physics, Abbe Center of Photonics, 
07745 Jena, Germany}
\author{Varun Raj Kaipalath}
\affiliation{Fraunhofer Institute for Applied Optics and Precision Engineering, 
07745 Jena, Germany}
\affiliation{Friedrich Schiller University Jena, Institute of Applied Physics, Abbe Center of Photonics, 
07745 Jena, Germany}
\author{Fabian Steinlechner}\email{fabian.steinlechner@uni-jena.de}
\affiliation{Fraunhofer Institute for Applied Optics and Precision Engineering, 
07745 Jena, Germany}
\affiliation{Friedrich Schiller University Jena, Institute of Applied Physics, Abbe Center of Photonics, 
07745 Jena, Germany}

\date{\today}

\begin{abstract}

Single-mode-fiber-coupled spontaneous parametric down-conversion (SPDC) sources are a key resource for photonic quantum technologies, but in single-crystal geometries brightness, heralding efficiency, and spectral purity remain constrained by intrinsic trade-offs. Here, we show how nonlinear interference in a cascaded two-crystal type-II SPDC source can be used to engineer the modal structure of SPDC emission, improving the brightness--heralding-efficiency trade-off by more than one order of magnitude beyond the single-crystal limit. We further demonstrate two routes to near-unity spectral purity while retaining high brightness and/or heralding efficiency, even with standard periodically poled crystals, and study the additional advantages of aperiodic poling with Gaussian phase matching. Using a spectrally resolved Laguerre--Gauss modal decomposition, we show that these improvements arise from mode-selective interference of spatial-spectral SPDC modes within the nonlinear interferometer. We experimentally validate the model through sum-frequency-generation measurements of the spatial-spectral state.
\end{abstract}

\maketitle


\section{Introduction}
\begin{figure}[t]
\centering
\begin{overpic}[unit=1mm,width=0.43\textwidth]{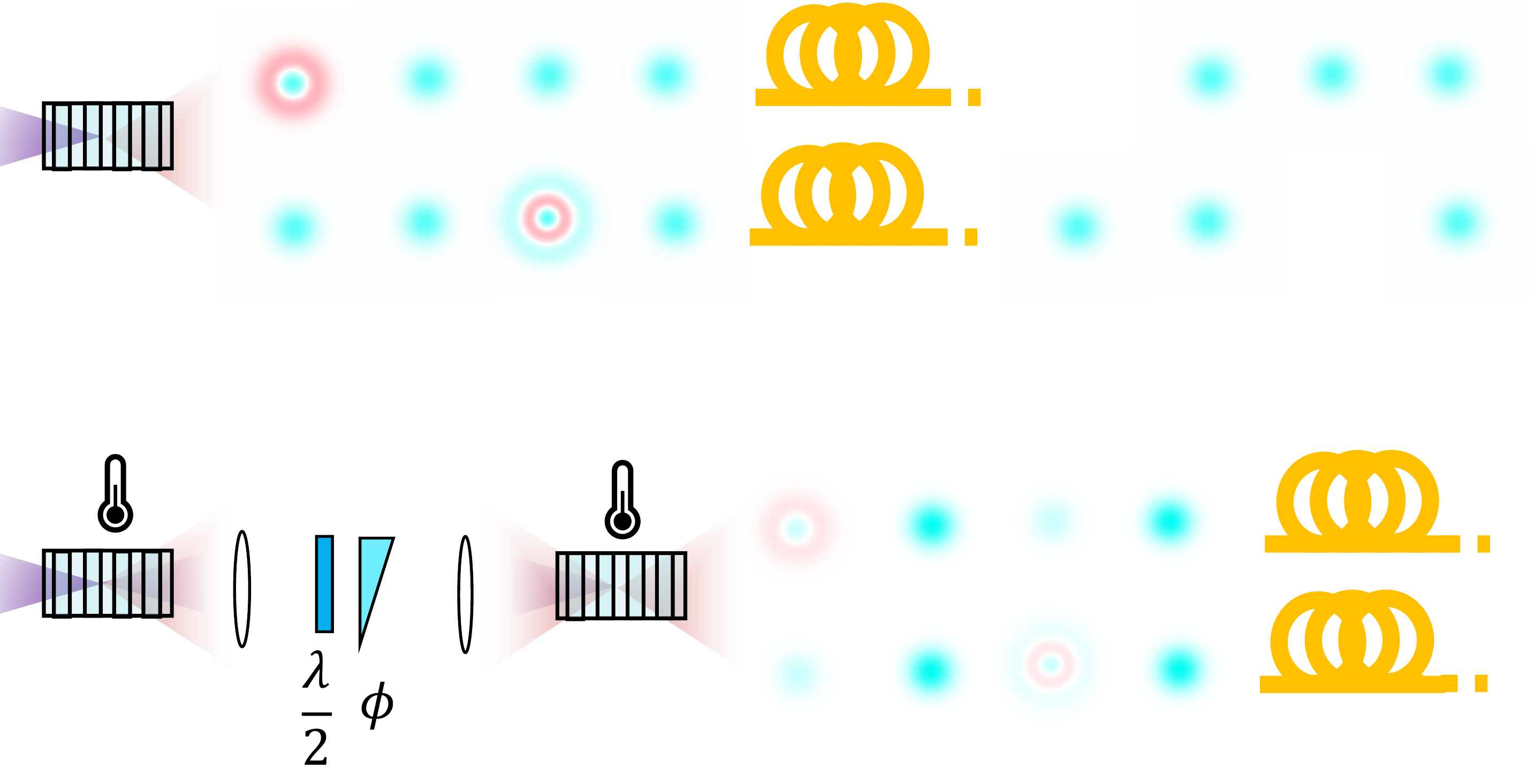}
	 \put(-4,36){\textbf{(a)}}	
	\put(-4,16){\textbf{(b)}}
\textbf{}
\end{overpic}
\caption{ (a) Conceptual illustration of the limitation of heralding efficiency in bulk SPDC with single-mode-fiber coupling: photon pairs emitted into mismatched spatial modes reduce the coincidence-to-singles ratio. (b) Schematic of the two-crystal configuration with intermediate polarization rotation ($\lambda/2$) and relative phase control $\phi$. The interference enhances generation into the joint Gaussian mode while partially suppressing higher-order mode contributions, resulting in increased brightness and heralding efficiency.}
\label{fig:concept_NLI}
\end{figure}

Spontaneous parametric down-conversion (SPDC) is a workhorse of quantum optics because of its versatility in generating heralded single photons\, \cite{helraded_single_photon_2016}, entangled photon pairs, and squeezed states of light\,\cite{Wu:1985_Squeezing_SPDC} with tailored correlations in space, time, and frequency\,\cite{Roux:2020_entanglement}. The development of high-performance single-mode-fiber (SMF)-coupled SPDC sources has enabled applications in quantum communication, quantum sensing, and photonic quantum computing. Two key performance metrics of such sources are the \textit{pair-collection probability} and the \textit{heralding efficiency}. The pair-collection probability quantifies the probability that both signal and idler photons are generated in the Gaussian mode supported by the fiber, and is therefore directly related to the source brightness. The heralding efficiency is the conditional probability that one photon is found in the collected mode given that its partner has been detected. Its limitation originates from the spatially multimode character of SPDC emission, illustrated in Fig.\,\ref{fig:concept_NLI}(a), where signal and idler photons can be emitted into different combinations of spatial modes. Since the SMF supports only the fundamental mode, photons generated in higher-order modes are lost, reducing the heralding efficiency. A third important parameter is the \textit{spectral purity}, which is related to the spectral correlations between the two photons and is essential for high-visibility multi-photon interference\,\cite{Pan2020,Pan2021,Jin2015SciRep,stefszky2025,dosSantosMartins:25,Pickston2023npjQI}.

Numerous works have studied the individual and joint optimization of these metrics for different phase-matching configurations by varying the focusing parameters of the pump, signal, and idler fields inside the nonlinear crystal (NLC)\,\cite{Kurtsiefer:01,Bovino:03,Dragan:04,Andrews:04,Ljunggren:05,Fedrizzi:07,Ling:08,Bennink:10,Palacios:11,Guerreiro:13,SteinlechnerPhD:2015,Minozzi:13}. These studies showed that tighter focusing generally enhances the pair-collection probability, whereas looser focusing improves the heralding efficiency, leading to a trade-off between the two. In addition, tight focusing modifies the spectral phase matching, typically increasing the bandwidth and introducing spectral asymmetry\,\cite{Varga:22_Bandwidth,Sevilla2025:Efficient}, which can reduce the spectral purity. The origin of this trade-off lies in the non-separability of the spatial and spectral degrees of freedom (DOF)\,\cite{Gatti_Xentanglement}, which becomes more pronounced under tight focusing\,\cite{Osorio2008,Guerreiro:13,Sevilla2025:Efficient}. As a result, spatial and spectral filtering alone are not an adequate solution, since they suppress unwanted modes only at the cost of additional loss and reduced brightness, particularly for pulsed pumping\,\cite{Meyer-Scott:2017_helrading_pulsed}.


Cascaded nonlinear processes have been used to tailor both spectral\,\cite{Uren_superlattices_206,Riazi2019Biphoton,Riazi2024Entangling,Ou_Li2020_3stage,Ou_Cui2022Programmable,Lemieux_Chechova2016_Freq_GVD,Bispectral_2018,Rui_2021,Prasad_2026} and spatial correlations\,\cite{Perez2014BrightSqueezed,Scharwald2025SchmidtOAM_sharapova,Paterova2020Superlattices} through nonlinear interference. Unlike direct filtering, nonlinear interference enables the selective enhancement or suppression of specific spatio-spectral modes at the generation stage, providing a lossless route to state engineering. Here, we study a cascaded two-crystal type-II SPDC source with intermediate polarization rotation and phase control, and show that nonlinear interference enhances Gaussian-mode coupling of spectrally uncorrelated photon pairs. As illustrated in Fig.\,\ref{fig:concept_NLI}(b), the second crystal coherently enhances joint emission into the Gaussian mode. This architecture has recently attracted attention because, when operated in the symmetry group-velocity-matching (SGVM) regime, it enables constructive or destructive interference over a broad bandwidth and can yield a fourfold brightness enhancement\,\cite{Lamas2001Stimulated,Pan2021}, making it attractive also for quantum sensing\,\cite{Ferreri2021spectrallymultimode,Qin2023_Unconditional}. In addition, Houde \textit{et al.}\,\cite{Houd2023_thebad} showed that this cascaded configuration can strongly suppress spectro-temporal distinguishability over a range of brightness levels.

Here, we leverage the nonlinear interference provided by this architecture to reveal further advantages in the context of efficient Gaussian-mode coupling of spectrally uncorrelated photon pairs. First, we show that high spectral purity can be achieved using standard periodically poled nonlinear crystals by destructively interfering the side lobes of the sinc-shaped phase-matching function through a combination of temperature tuning and phase control. In particular, we focus on the SGVM regime, which occurs in type-II KTP crystals in the C-band wavelength range\,\cite{Ferreri2021spectrallymultimode,Ansari:18}. In this configuration, we show that near-unity heralding efficiency can be achieved together with a twofold brightness enhancement with respect to the single-crystal case. More generally, we find that the brightness-heralding trade-off is improved by more than one order of magnitude; that is, for a fixed heralding efficiency, the cascaded source can be more than ten times brighter than its single-crystal counterpart. Using a spectrally resolved Laguerre-Gauss mode decomposition of the spatio-spectral SPDC state, we show that this enhancement originates from the way different spatio-spectral modes interfere within the nonlinear interferometer. Furthermore, we show that near-unity spectral purity ($\approx0.99$) remains achievable even under tight focusing while maintaining close-to-optimal brightness and high heralding efficiency ($>0.9$). We extend the analysis to aperiodically poled crystals, for which we observe similar advantages and find that the spectral purity is not substantially degraded under tight focusing, in contrast to the single-crystal configuration. Finally, we experimentally validate our model by using sum-frequency generation (SFG) measurements to measure the spectral correlations associated with different spatial modes\,\cite{Karspinski_2009,Kaneda:20,Helt:15}.

The remainder of the paper is organized as follows. In Section~\ref{sec:stimulated_SPDC}, we illustrate the nonlinear interferometer in the simplified plane-wave approximation and introduce the definition of spectral purity. In Section~\ref{sec:spatial-spectral_correlation} we extend the description incorporating multiple spatial modes and spatial-spectral correlations and introduce the definitions of pair-collection probability and heralding efficiency. Section\,\ref{sec:methods} details the numerical parameters used in our simulations and further considerations. In Section~\ref{sec:results}, we show the main results of our study. In Section\,\ref{sec:experiment}, we show experimental validation of some of our theoretical results. Finally, we summarize our conclusions in Section~\ref{sec:discussion}.

\section{Brightness and spectral purity enhancement}\label{sec:stimulated_SPDC}
Here we illustrate the concept of brightness enhancement and phase matching engineering using the particular configuration of a polarization rotation for the SPDC photons and phase control. For now we restrict to the case of plane waves, therefore, only taking into account the spectral DOF.  

The phase matching function (PMF) at the center of each individual crystal is given by 
\begin{equation}\label{eq:PM_1}
F(\omega_\mathrm{s},\omega_\mathrm{i}) = \sinc\Big(\frac{\Delta k(\omega_\mathrm{s},\omega_\mathrm{i}) L}{2} \Big)     
\end{equation}

Here, $\Delta k(\omega_\mathrm{s},\omega_\mathrm{i})= k_\mathrm{p}(\omega_\mathrm{s}+\omega_\mathrm{i})-k_\mathrm{s}(\omega_\mathrm{s})-k_\mathrm{i}(\omega_\mathrm{i})+2\pi/\Lambda$ is the longitudinal phase mismatch, with $\Lambda$ being the period of the poling. Taking into account the first order dispersion term, this can be expressed as

\begin{equation}\label{eq:dK_1}
 \Delta k(\Omega_\mathrm{s},\Omega_\mathrm{i})=\frac{2\phi_T}{L}+(v_\mathrm{p}^{-1}-v_\mathrm{s}^{-1})\Omega_\mathrm{s}+(v_\mathrm{p}^{-1}-v_\mathrm{i}^{-1})\Omega_\mathrm{i}
\end{equation}


Here, $v_\mathrm{j}$ denotes the group velocity and $\Omega_\mathrm{j}$ the frequency detuning from the central frequency $\omega^0_\mathrm{j}$, such that $\omega_\mathrm{j}=\omega^0_\mathrm{j}+\Omega_\mathrm{j}$, with $\mathrm{j}=\mathrm{p},\mathrm{s},\mathrm{i}$. Furthermore, $\Delta k_0=k_{\mathrm{p},0}-k_{\mathrm{s},0}-k_{\mathrm{i},0}+2\pi/\Lambda$ is the longitudinal phase mismatch at the central frequencies. For a given initial configuration, this quantity is typically zero, but it can be shifted away from zero by temperature tuning, mainly through the thermo-optic dependence of the refractive indices and, to a lesser extent, through thermal expansion of the poling period\,\cite{Nina_cryo_2023,Lerch:13_tuningT0,Song:25_temp_pump}. We therefore describe this contribution by the temperature-dependent phase-matching offset $\phi_T=\Delta k_0L/2$, which is a dimensionless phase expressed in radians.

The polarization rotation performed by the half-wave plate (HWP) flips the labels between signal and idler photons, i.e., it exchanges $\Omega_\mathrm{s}\leftrightarrow\Omega_\mathrm{i}$. We denote this flipping operation by $[x]_{f}$. Note that the polarization of the pump field is not affected, i.e. the wave plate acts as a full-wave plate (FWP) for the pump wavelength, which can be readily achieved in experiment\,\cite{Pan2021}. The resulting phase-matching function for the two crystal configuration can be written up to a global phase as

\begin{equation}\label{eq:NLI_1}
    \Phi(\Omega_\mathrm{s},\Omega_\mathrm{i})=[F(\Omega_\mathrm{s},\Omega_\mathrm{i})e^{-i\frac{\Delta k L}{2}}]_{f}e^{i(-\frac{\Delta k L}{2}+\phi)}+F(\Omega_\mathrm{s},\Omega_\mathrm{i}).
\end{equation}

Here, the term $\Delta k L$ in the exponential functions,  accounts for the propagation of signal and idler photons through half of the first crystal and the second crystal, before and after flipping operation, respectively, and the propagation of the pump from the half of the first crystal to the half of the second crystal. Additionally, we consider a controllable constant phase $\phi=\phi_\mathrm{p}-\phi_\mathrm{s}-\phi_\mathrm{i}$. In the considered common-path configuration, this phase can be set, e.g using a thin birefringent crystal such as a Soleil-Babinet compensator. The term $[F(\Omega_\mathrm{s},\Omega_\mathrm{i})]_{f}$ can be written the same as Eq.\,\ref{eq:PM_1} by replacing $\Delta k$ by $\Delta k'$, where 

\begin{equation}\label{eq:dK_1disp}
    \Delta k'(\Omega_\mathrm{s},\Omega_\mathrm{i})=\frac{2\phi_T}{L}+(v_\mathrm{p}^{-1}-v_\mathrm{s}^{-1})\Omega_\mathrm{i}+(v_\mathrm{p}^{-1}-v_\mathrm{i}^{-1})\Omega_\mathrm{s}.
\end{equation}


When operating in the SGVM regime, $2 v_\mathrm{p}^{-1}=v_\mathrm{i}^{-1}+v_\mathrm{s}^{-1}$, it is easy to see that  
\begin{equation}
\Delta k(\Omega)=\frac{2\phi_T}{L}+D\Omega,
\qquad
\Delta k'(\Omega)=\frac{2\phi_T}{L}-D\Omega,
\end{equation}
where $\Omega=(\Omega_\mathrm{i}-\Omega_\mathrm{s})/2$ and $D=v^{-1}_\mathrm{s}-v^{-1}_\mathrm{i}$. Note that these expressions take the same single-detuning form for a monochromatic pump, where  $\Omega_\mathrm{s}=-\Omega_\mathrm{i}$. Therefore, our results with regard to brightness and heralding efficiency apply to the continuous-wave pump even outside the SGVM regime. 

Interestingly, for the particular case of  $\phi_T=0$, $\Delta k'=\Delta k$ and Eq.\,\ref{eq:NLI_1} can be written as (up to a global phase)

\begin{equation}\label{eq:NLI_1_compact}
\Phi(\Omega) = (1+e^{i\phi})
\sinc\Big(\frac{\Delta k L}{2}\Big)\,,
\end{equation}


\begin{figure}[t]
\centering
\begin{overpic}[unit=1mm,width=0.47\textwidth]{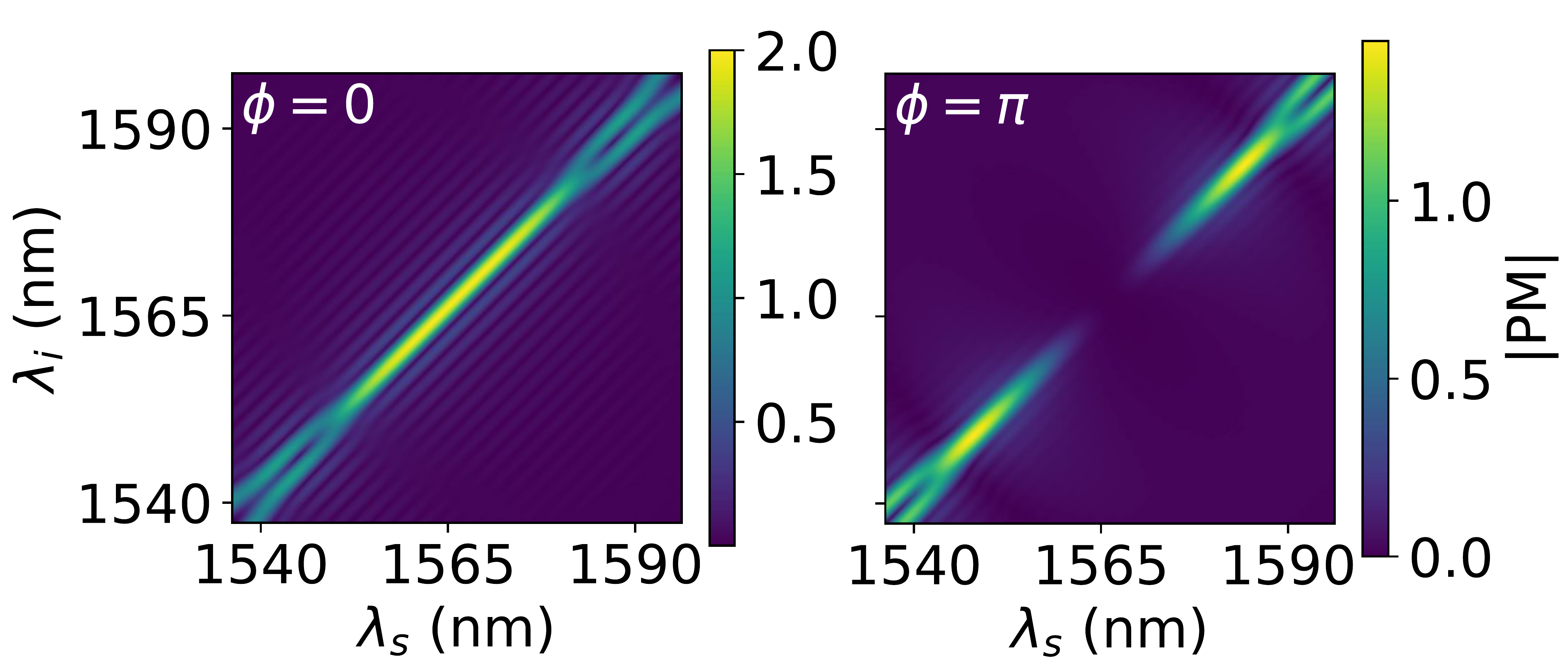}
\textbf{}
\end{overpic}
\caption{
Effective phase matching under the plane-wave approximation for constructive interference ($\phi=0$) (left) and destructive interference ($\phi=\pi$) (right). The plots show that the interference range is limited by higher-order dispersion terms.}
\label{fig:PMF_planewave}
\end{figure}

When setting $\phi=0\,(\pi)$, we can obtain constructive (destructive) interference over broad bandwidth (limited by the higher-order dispersion terms), obtaining an enhancement of the photon flux by a factor of 4. The resulting PMF is depicted in Fig.\,\ref{fig:PMF_planewave} for the case of KTP at around $1560\,nm$ for constructive (left) and destructive (right) interference. From the figures the effect of the flip operation can be observed. The PMF from the first crystal is reflected around the $\Omega_\mathrm{s}+ \Omega_\mathrm{i}$-axis, resulting in a X-shaped PMF for two crystals.

Due to the $45\degree$ orientation of the PMF, spectrally uncorrelated photon sources can be realized. The spectral purity can be calculated from the spectral correlations in the the \textit{joint spectral amplitude} (JSA). The JSA is the product of the pump spectral function $f(\Omega_\mathrm{p})$ and the phase matching function $J(\Omega_\mathrm{s},\Omega_\mathrm{i}) = f(\Omega_\mathrm{p})\times\Phi(\Omega_\mathrm{s},\Omega_\mathrm{i})$, where $\Omega_\mathrm{p}=\Omega_\mathrm{s}+\Omega_\mathrm{i}$, due to energy conservation. Performing singular-value decomposition (SVD) on the JSA, the Schmidt coefficients ($\sqrt{\lambda_k}$) can be extracted\,\cite{Law_2000,Grice_2001}. Finally, the purity is given by $\mathcal{P}=1/\sum_k\lambda_k^2$\,\cite{Uren2005PureSinglePhoton,Mosley_2008}. Considering a sinc-shaped phase matching (periodically poled crystal) under SGVM and a spectrally Gaussian pump, $G(\Omega_\mathrm{p})=\exp{(-\frac{\Omega_\mathrm{p}^2}{\sigma_\mathrm{p}^2}})$, of bandwidth $\sigma_p$, the maximum spectral purity is $\mathcal{P}\approx0.81$.

\begin{figure}[t]
\centering
\begin{overpic}[unit=1mm,width=0.48\textwidth]{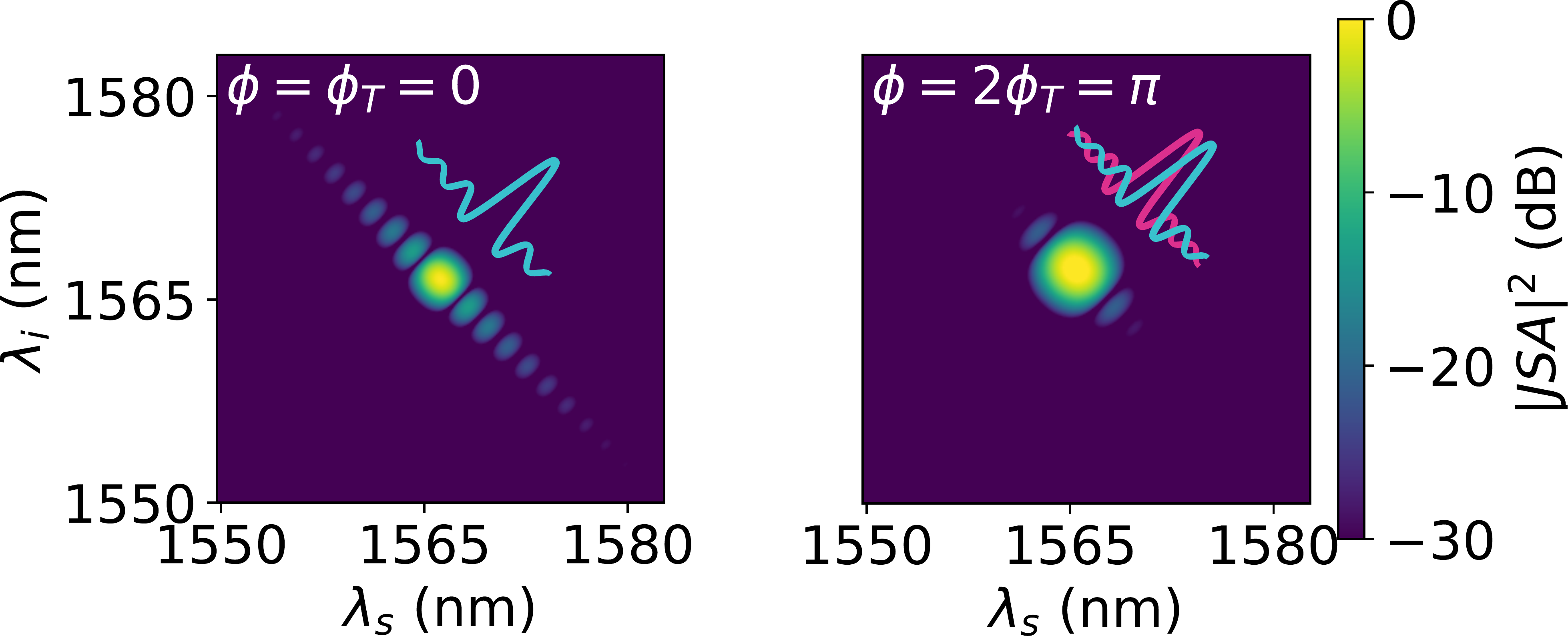}
	\put(-2,32){\textbf{(a)}}
    \put(40,32){\textbf{(b)}}
\textbf{}
\end{overpic}
\caption{
(a) Joint spectral intensity ($|JSA|^2$) for the zero-detuning configuration, $\phi=\phi_T=0$, and (b) for the detuned configuration, $\phi=2\phi_T=\pi$. The latter exhibits strong sidelobe suppression, increasing the spectral purity from $\approx 0.81$ to $\approx 0.98$.}
\label{fig:JSA_planewave}
\end{figure}

We next show that, in the proposed cascaded configuration, the temperature dependence can be leveraged to increase the spectral purity. Up to a global phase, Eq.~\ref{eq:NLI_1} can be written as

\begin{equation}\label{eq:NLI_1_phi_t}
\Phi(\Omega) =
\sinc\Big(\frac{\Delta k L}{2}+\phi_T\Big)+\sinc\Big(\frac{\Delta k L}{2}-\phi_T\Big) e^{i(\phi-2\phi_T)}\,.
\end{equation}
Here we have taken the temperature dependence $\phi_T$ outside of the definition of $\Delta k$ for clearer visualization.
By setting $\phi_T=\pi/2$ and $\phi=2\phi_T$, the negative sidelobes of the phase-matching function (PMF) from the first crystal overlap with the positive sidelobes from the second crystal (and vice versa), leading to strong sidelobe suppression while enhancing the central peak. The corresponding joint spectral intensities (JSIs) for a single crystal and for the optimized two-crystal configuration are shown in Fig.~\ref{fig:JSA_planewave}(a) and (b), respectively. The data are plotted in dB to better visualize the sidelobe reduction. For this configuration, the spectral purity can be increased to $\mathcal{P}=0.98$ while still achieving a photon-flux enhancement of approximately a factor of two. The pump bandwidth must be optimized accordingly to account for the increased PMF bandwidth. 



\section{Spatial-Spectral correlations in SPDC}\label{sec:spatial-spectral_correlation}
Now we introduce the effect of considering fields with finite transverse amplitude. 
In this case, the phase matching function also depends on the transverse momentum ($\bm{q_\mathrm{j}}$) of signal and idler photons through $k^z_\mathrm{j}=\sqrt{k^2_\mathrm{j}-|\bm{q_\mathrm{j}}|^2}$. The longitudinal phase mismatch in the paraxial approximation ($k_\mathrm{j}\gg|\bm{q_\mathrm{j}}|$) is given by\,\cite{Baghdasaryan2022:Generalized}


\begin{align}\label{eq:Dk_paraxial}
 \Delta k_{S}=\frac{2\phi_T}{L}+
 D\Omega+\frac{|\bm{q_\mathrm{p}}|^2}{2k_\mathrm{p}(\omega^0_\mathrm{p})}-\frac{|\bm{q_\mathrm{s}}|^2}{2k_\mathrm{s}(\omega^0_\mathrm{s})}-\frac{|\bm{q_\mathrm{i}}|^2}{2k_\mathrm{i}(\omega^0_\mathrm{i})}
\end{align}

Here we added the subscript $S$ to distinguish it from Eq.\,\ref{eq:dK_1}. The spatial domain can be expressed in different bases. Laguerre-Gauss (LG) basis is typically used to describe SPDC. It is a discrete basis which is characterized by two indices: the radial $p$ and the azimuthal $\ell$ indices, the latter related to the orbital angular momentum (OAM)\,\cite{Mair2001OAMEntanglement}. The decomposition of the spatial correlations in the LG basis is particularly interesting as the OAM is conserved in collinear SPDC ($\ell_\mathrm{p}=\ell_\mathrm{i}+\ell_\mathrm{s}$). 

In this basis, the PMF at the center of the crystal for a particular mode combination of pump, signal and idler is given by\,\cite{Sevilla2024}

\begin{align}\label{coe1}
    C^{\ell_\mathrm{p},\ell_\mathrm{s},\ell_\mathrm{i}}_{p_\mathrm{p}, p_\mathrm{s}, p_\mathrm{i}}(\Omega) 
    =    \frac{1}{\sqrt{N}}\iint  d\bm{q}_\mathrm{s} \: d\bm{q}_\mathrm{i}\int^{L/2}_{-L/2}dz
    \,\chi^{(2)}(z) e^{(-i\Delta k_S z)}\nonumber&\\
    \times\mathrm{LG}_{p_\mathrm{p}}^{\ell_\mathrm{p}}(\bm{q}_\mathrm{p})  
     [\mathrm{LG}_{p_\mathrm{s}}^{\ell_\mathrm{s}}(\bm{q}_\mathrm{s})]^*         [\mathrm{LG}_{p_\mathrm{i}}^{\ell_\mathrm{i}}(\bm{q}_\mathrm{i})]^*\,.
\end{align}

Here, $N$ is a normalization factor considering the generation of zero (vacuum) or one photon pair in any spatial-spectral state. This way $|C^{\ell_\mathrm{p},\ell_\mathrm{s},\ell_\mathrm{i}}_{p_\mathrm{p}, p_\mathrm{s}, p_\mathrm{i}}(\Omega)|^2$ is related to the probability of generating a photon pair in a given spatial-spectral mode per pump photon in a single crystal. Moreover, we leave the $z-$dependency of the nonlinearity $\chi^{(2)}(z)$, as we will consider the cases of periodic and aperiodic poling. 



Unlike the plane-wave discussion in the previous section, $C^{\ell_\mathrm{p},\ell_\mathrm{s},\ell_\mathrm{i}}_{p_\mathrm{p}, p_\mathrm{s}, p_\mathrm{i}}(\Omega)$ transforms upon propagation. For simplicity, in the following derivations we consider an ideal 4-$f$ system that images the plane at the center of the first crystal onto the center of the second crystal. For a more general calculation, we refer the reader to the Appendix, where we use the ABCD matrix formalism for LG beams\,\cite{Tache:87,Vallone:17} to propagate the pump mode and biphoton state through a more realistic optical system.

As a further consideration, we will use a fundamental Gaussian beam for the pump ($p_\mathrm{p}=\ell_\mathrm{p}=0$). Due to OAM conservation, modes where $|\ell|>0$ can be neglected, since we are only interested in the scenarios in which at least one SPDC photon is in a Gaussian mode. Hereafter, these indices will be omitted for brevity, leaving only dependency on the radial indices.

The resulting PMF is given by


\begin{align}\label{eq:NLI_2}
    \Phi_{p_\mathrm{s}, p_\mathrm{i}}(\Omega)=[C_{p_\mathrm{s}, p_\mathrm{i}}(\Omega)e^{-i\frac{\Delta k L}{2}}]_{f}
    +C_{p_\mathrm{s}, p_\mathrm{i}}(\Omega)\,e^{i\beta(\Omega_\mathrm{s},\Omega_\mathrm{i})}.
\end{align}

Here, $\beta(\Omega_\mathrm{s},\Omega_\mathrm{i})=\frac{\Delta k(\Omega) L}{2}+H(\Omega_\mathrm{s},\Omega_\mathrm{i})+\phi$ accounts for all the dispersion seen by the pump, signal and idler photons. In particular $H(\Omega_\mathrm{s},\Omega_\mathrm{i})$ takes into account additional dispersion, e.g. due to the lenses in the 4-f system. Note that the flip operation also exchanges the spatial mode label $p_\mathrm{s}\xleftrightarrow{}p_\mathrm{i}$. In general, $C_{x,0}\neq C_{0,x}$ due to the small birefringence; however, the difference is nearly negligible. In previous work, it has been shown how the spectrum of different joint spatial modes differ and become more asymmetric and shifted from the central frequency, especially as the focusing becomes stronger\,\cite{Sevilla2024,Sevilla2025:Efficient}. This means that the PMF is not symmetric with respect to the frequency labels $C_{x,y}(\Omega)\neq C_{x,y}(-\Omega)$.


The probability of generation of a pair in a joint spatial mode is given by

\begin{equation}
    S_{p_\mathrm{s}, p_\mathrm{i}}=\int d\Omega_\mathrm{s}d\Omega_\mathrm{i} \,|f(\Omega_\mathrm{p})\Phi_{p_\mathrm{s}, p_\mathrm{i}}(\Omega_\mathrm{s},\Omega_\mathrm{i})|^2,
\end{equation}

where $f(\Omega_\mathrm{p})$ is the spectral mode of the pump beam, $\Omega_\mathrm{p}=\Omega_\mathrm{s}+\Omega_\mathrm{i}$ is the frequency detuning of the pump. Note that for the single-crystal configuration $\Phi_{p_\mathrm{s}, p_\mathrm{i}}$ is replaced by $C_{p_\mathrm{s}, p_\mathrm{i}}$.


Based on this, we define the pair-collection probability as the probability that both the signal and idler photons are generated in Gaussian modes and are therefore coupled into the SMF. This quantity is given by $S_{0,0}$.

The heralding efficiency is calculated from the ratio of the probability that signal and idler modes are in the Gaussian mode, and the probability of only one being a Gaussian by tracing over the spatial state of its partner 

\begin{equation}\label{eq:H_sum}
    H=  \frac{S_{0,0}}{\sum_{p} S_{0,p}}\,. 
\end{equation}



For a single crystal, different heralding efficiencies need to be defined for signal and idler photons, $H_\mathrm{s}$ and $H_\mathrm{i}$, respectively,

\begin{equation}\label{eq:decomposition_gauss}
    H_\mathrm{s} =  \frac{S_{0,0}}{\sum_{p_\mathrm{i}} S_{0,p_\mathrm{i}}}    , \,\,\, H_\mathrm{i} =  \frac{S_{0,0}}{\sum_{p_\mathrm{s}} S_{p_\mathrm{s},0}}    \,. 
\end{equation}

For simplicity, the so-called \textit{symmetric heralding efficiency} is typically used, which is defined as $H_{sym}=\sqrt{H_\mathrm{s}H_\mathrm{i}}$\,\cite{Bennink:10}. However, in the near-degenerate case $H_{sym}\approx H_{\mathrm{s}} \approx H_{\mathrm{i}}$\,\cite{Sevilla2025:Efficient}.

The pair-collection probability and heralding efficiency are dependent on the waist parameters of the pump ($w_\mathrm{p}$), signal ($w_\mathrm{s}$), and idler ($w_\mathrm{i}$), and therefore can be optimized over them. However, it is more useful to define the dimensionless focal parameter $\xi_j=\frac{L\lambda_\mathrm{j}}{2\pi n_jw_j^2}$ instead of the waist parameter, as it makes the results applicable to any crystal length and wavelength operation\,\cite{Sevilla2025:Efficient}. Here, $L$ is the length of the crystal, $\lambda$ is the wavelength and $n$ is the refractive index. 
In the following section, we first consider the single-crystal configuration as a benchmark and then compare it with the cascaded two-crystal architecture.

\section{Methods}\label{sec:methods}

The figures of merit considered in this work---pair-collection probability, heralding efficiency, and spectral purity---were introduced in the preceding sections together with their dependence on the focusing conditions, temperature, and the relative phase between the three fields in the nonlinear interferometer. Here, we summarize only the assumptions and conventions relevant for the present study, following our previous analysis in Ref.~\cite{Sevilla2025:Efficient}. 

To compare source performance across these optimization parameters, we normalize the collected pair probability to the optimum value obtained for the single-crystal source and define the corresponding \textit{relative brightness} as
\begin{equation}
B=
\frac{S_{0,0}}{\max_{\xi_\mathrm{p},\xi_\mathrm{s},\xi_\mathrm{i}} S_{0,0}^{sc}}\, .
\end{equation}

Throughout the calculations, we impose $\xi_\mathrm{s}=\xi_\mathrm{i}$. This is well justified for the single-crystal case, where the spatial correlations remain close to symmetric despite the weak birefringence in the type-II nonlinear process\,\cite{Sevilla2025:Efficient}. In the cascaded configuration, the same approximation remains appropriate because the effective signal contribution contains the idler mode generated in the first crystal together with the signal mode generated in the second crystal. Similarly for the effective idler contribution. We further limit the analysis to focusing parameters $\xi\leq 10$, which covers the regime relevant for typical experiments.

Since our main interest is in high-heralding-efficiency operation, we focus on the region $H\ge0.7$. In this regime, truncating the radial-mode expansion at $p_\mathrm{j}=4$ is sufficient, as higher-order contributions change $H$ by less than $0.02$ at this threshold and become quickly negligible for larger values of $H$.

 Whenever possible, the results are expressed in terms of the dimensionless variables $\xi$ and $D\Omega L$ in order to facilitate comparison with other experimental implementations. All quantities reported below are obtained from numerical simulations and the ranges and step sizes for $\xi$ and $D\Omega L$ are listed in the Appendix.

In the following section, we first consider the single-crystal configuration as a benchmark and then compare it with the cascaded two-crystal architecture.

\section{Results}\label{sec:results}

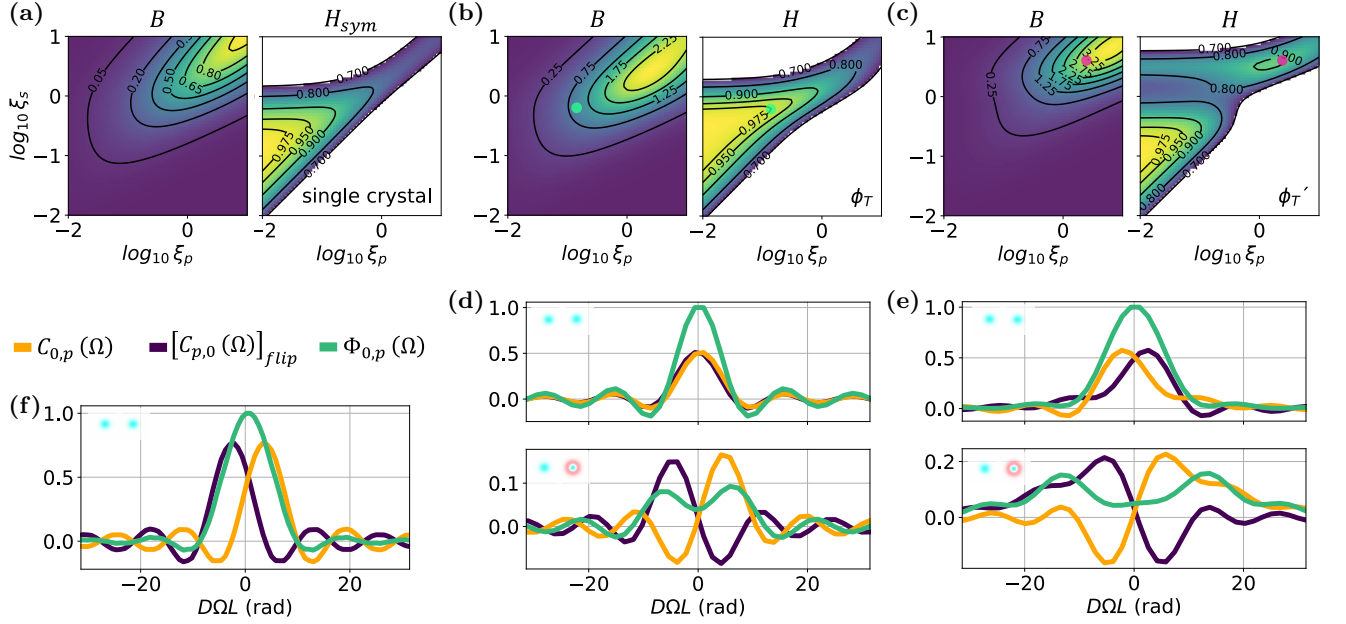
\begin{figure*}[t]
\centering
\begin{overpic}[unit=1mm,width=1.0\textwidth]{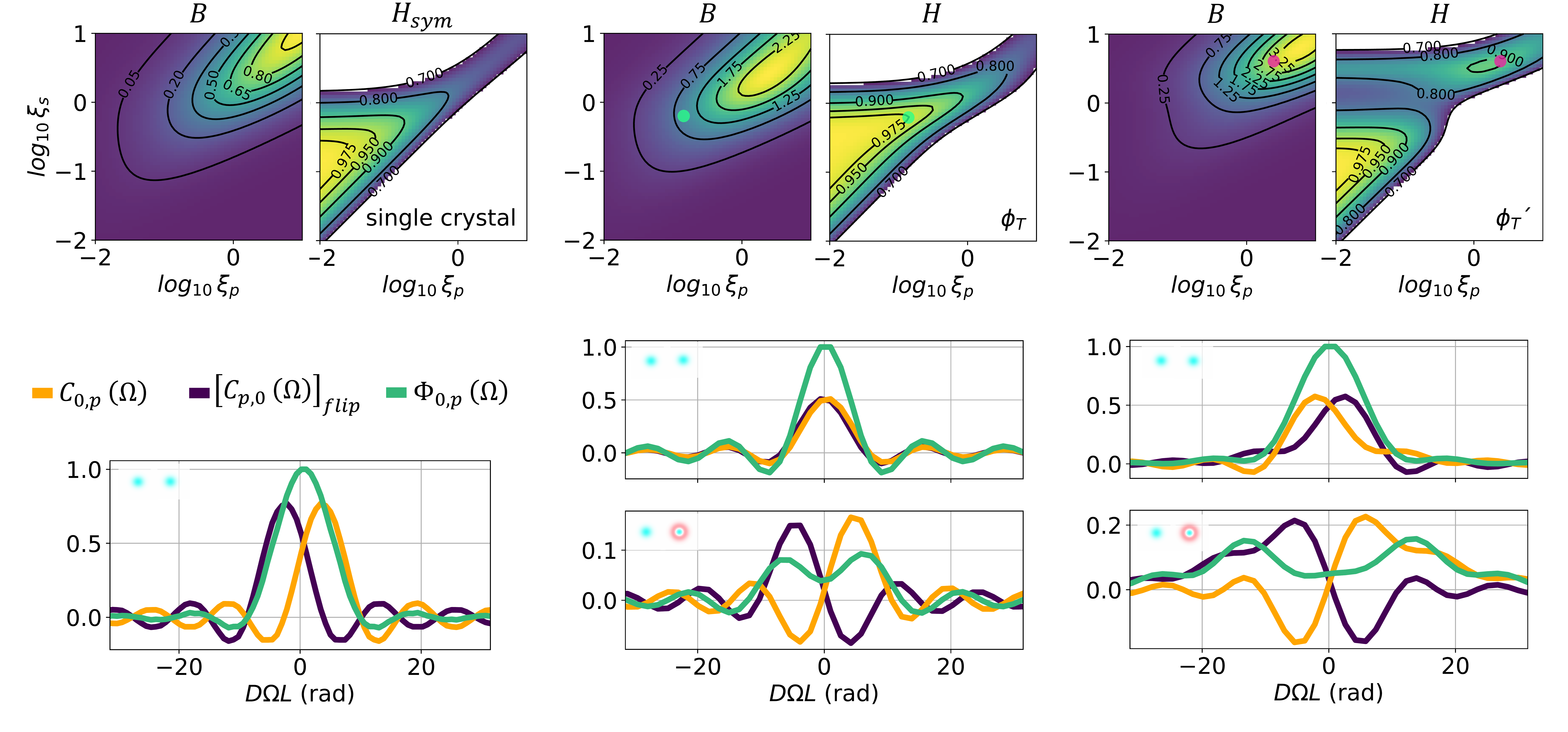}
	 \put(3,82){\textbf{(a)}}	
     \put(61,82){\textbf{(b)}}	
      \put(119,82){\textbf{(c)}}
	    \put(61,44){\textbf{(d)}}
 	     \put(119,44){\textbf{(e)}}
           \put(3,30){\textbf{(f)}}

\end{overpic}
\caption{
Relative brightness (left) and heralding efficiency (right) for the (a) single-crystal configuration and the two-crystal configuration with phase-mismatch settings (b) $\phi_T$ and (c) $\phi_T'$. (d),(e) Spectral amplitudes for the joint spatial modes $p_\mathrm{s}=p_\mathrm{i}=0$ (top) and $p_\mathrm{s}=0$, $p_\mathrm{i}=1$ (bottom) at the operating points marked in green and magenta in (b) and (c), respectively. Blue and yellow lines denote the contribution from each crystal, and the green curve the effective spectral amplitude after nonlinear interference. The interference enhances the Gaussian mode while partially suppressing higher-order contributions, improving the brightness--heralding-efficiency trade-off. In (e), sidelobe suppression yields a spectral purity of $P>0.99$. (f) Spectral amplitudes in the loose-focusing regime with high heralding efficiency and optimized spectral purity via temperature tuning.}
\label{fig:B_H_Spect}
\end{figure*}

Figure\,\ref{fig:B_H_Spect}(a) shows $B$ and $H_\mathrm{sym}$ for the single-crystal configuration as functions of $\xi_\mathrm{p}$ and $\xi_\mathrm{s}$. Comparison of the two maps shows that $B$ is maximized under tighter focusing, whereas $H_\mathrm{sym}$ is favored by looser focusing, leading to the well-known trade-off between brightness and heralding efficiency\,\cite{Bennink:10}.

The severity of this trade-off is evident from the contour lines. To achieve $H_\mathrm{sym}>0.90$, the brightness must be reduced by approximately a factor of 5, corresponding to $B<0.22$. Reaching $H_\mathrm{sym}>0.97$ requires an even stronger reduction, with $B<0.08$, i.e., more than one order of magnitude below its maximum value.

We now show how nonlinear interference substantially improves this trade-off. Figures\,\ref{fig:B_H_Spect}(b) and \ref{fig:B_H_Spect}(c) show $B$ and $H$ for the two-crystal configuration for two different values of $\phi_T$, with $\phi=2\phi_T$. As expected, $B$ can exceed unity because it is normalized to the maximum value attainable in the single-crystal configuration, allowing direct comparison between the two schemes. For both choices of $\phi_T$, $B$ exhibits a well-defined maximum, in contrast to the single-crystal case. Moreover, as indicated by the contour lines, the high-heralding-efficiency region extends further toward tighter focusing, approaching the region of optimal brightness. This directly reflects an improved trade-off between brightness and heralding efficiency.

We emphasize that this improvement cannot be reproduced simply by using a crystal of twice the length. For collinear type-II quasi-phase matching in bulk crystals, both $B$ and $H_\mathrm{sym}$ are independent of $L$, since the additional dispersion and diffraction cancels out the benefit of a longer interaction length~\cite{Sevilla2025:Efficient}. In the two-crystal configuration, however, these effects are effectively compensated: the polarization rotation exchanges the signal and idler labels, thereby compensating the extra dispersion, while the 4-$f$ system mitigates the effect of free-space diffraction between the two crystals.

In Fig.\,\ref{fig:B_H_Spect}(b), the green markers identify a point at $(\xi_\mathrm{s}=0.71,\,\xi_\mathrm{p}=0.16)$ where $B=1$ and $H\approx0.97$ are achieved simultaneously. Compared with the single-crystal configuration, this corresponds to an approximately 12-fold enhancement in $B$ at the same heralding efficiency.

To understand the origin of this enhancement, Fig.\,\ref{fig:B_H_Spect}(d) shows the spectral amplitudes for the joint spatial modes $p_\mathrm{s}=p_\mathrm{i}=0$ (top) and $p_\mathrm{s}=0$, $p_\mathrm{i}=1$ (bottom). In each panel, the contributions from the first crystal, $C_{0,p}(\Omega)$ (yellow), second crystal, $[C_{p,0}(\Omega)]_f$ (blue), and the resulting interfered amplitude, $\Phi_{0,p}(\Omega)$ (green), are plotted. For the Gaussian mode ($p_\mathrm{s}=p_\mathrm{i}=0$), the independent contributions are symmetric about the degenerate frequency and overlap nearly perfectly, giving rise to fully constructive interference and a fourfold increase in generation probability. However, for the higher-order mode ($p_\mathrm{s}=0$, $p_\mathrm{i}=1$), each amplitude contains positive and negative contributions near $\Omega=0$. After the flip operation, the two crystal contributions overlap with opposite sign, leading to partial suppression of this mode. Together, the enhancement of the Gaussian mode and the suppression of the higher-order mode increase the heralding efficiency to $H\approx0.97$.

We now examine the effect of varying the phase mismatch $\phi_T$. Figure\,\ref{fig:B_H_Spect}(c) shows that increasing $\phi_T$ shifts the maximum brightness to tighter focusing conditions, while reducing $B$ under looser focusing. This behavior originates from the spectral shift induced by tight focusing\,\cite{Boyd:68,Bennink:10,Sevilla2025:Efficient}. This offset reduces the interference, but can be partially compensated by tuning $\phi_T$. Another notable feature is that the heralding efficiency improves in the region of highest brightness, as indicated by the $H=0.9$ contour. For these reasons, $\phi_T$ must be included as an optimization parameter.

The magenta markers in Fig.\,\ref{fig:B_H_Spect}(c) identify a particularly attractive operating point at $\xi_\mathrm{p}=2.24$ and $\xi_\mathrm{s}=3.55$, where $B=3$ and $H=0.9$ are achieved. Relative to the single-crystal configuration ($H_{sym}=0.9$, $B=0.22$), this corresponds to more than 13-fold enhancement in $B$ at the same heralding efficiency. Interestingly, this operating point also exhibits favorable spectral properties. The corresponding spectral amplitudes for the same two modes are shown in Fig.\,\ref{fig:B_H_Spect}(e). In this case, the effect of tighter focusing on the spectral amplitudes is evident: the spectra broaden and become asymmetric, reducing their overlap and thereby lowering the enhancement of the Gaussian mode $p_\mathrm{s}=p_\mathrm{i}=0$ to $\approx3.57$ instead of 4. Nevertheless, the resulting amplitude $\Phi_{0,0}(\Omega)$ is nearly symmetric, substantially reducing the spectral distinguishability between the two photons. Most notably, the sidelobes are strongly suppressed, increasing the achievable spectral purity to $\mathcal{P}>0.99$. By comparison, in the single-crystal case the maximum attainable spectral purity is only $P\approx0.93$, obtained at $B\approx0.8$ and $H_\mathrm{sym}\approx0.76$, despite the reduced spectral distinguishability associated with the asymmetric spectral amplitude.


Although high spectral purity is readily achieved under tight focusing, it can also be realized in the loose-focusing regime, where high heralding efficiency is obtained, using the method introduced in Section\,\ref{sec:stimulated_SPDC}. To this end, we fix the focusing parameters to the single-crystal operating point that provides the target heralding efficiency and then tune $\phi_T$ to maximize sidelobe suppression. The result is shown in Fig.\,\ref{fig:B_H_Spect}(f). For the reference case $H_\mathrm{sym}=0.967$, we obtain $H=0.975$, a spectral purity of $\mathcal{P}=0.972$, and a twofold enhancement in pair-collection probability, corresponding to $B=0.145$. 


\subsection{Trade-off comparison using periodically-poled crystals } \label{subsec:Tradeoff_PP}

\begin{figure}[t]
\centering
\begin{overpic}[unit=1mm,width=0.42\textwidth]{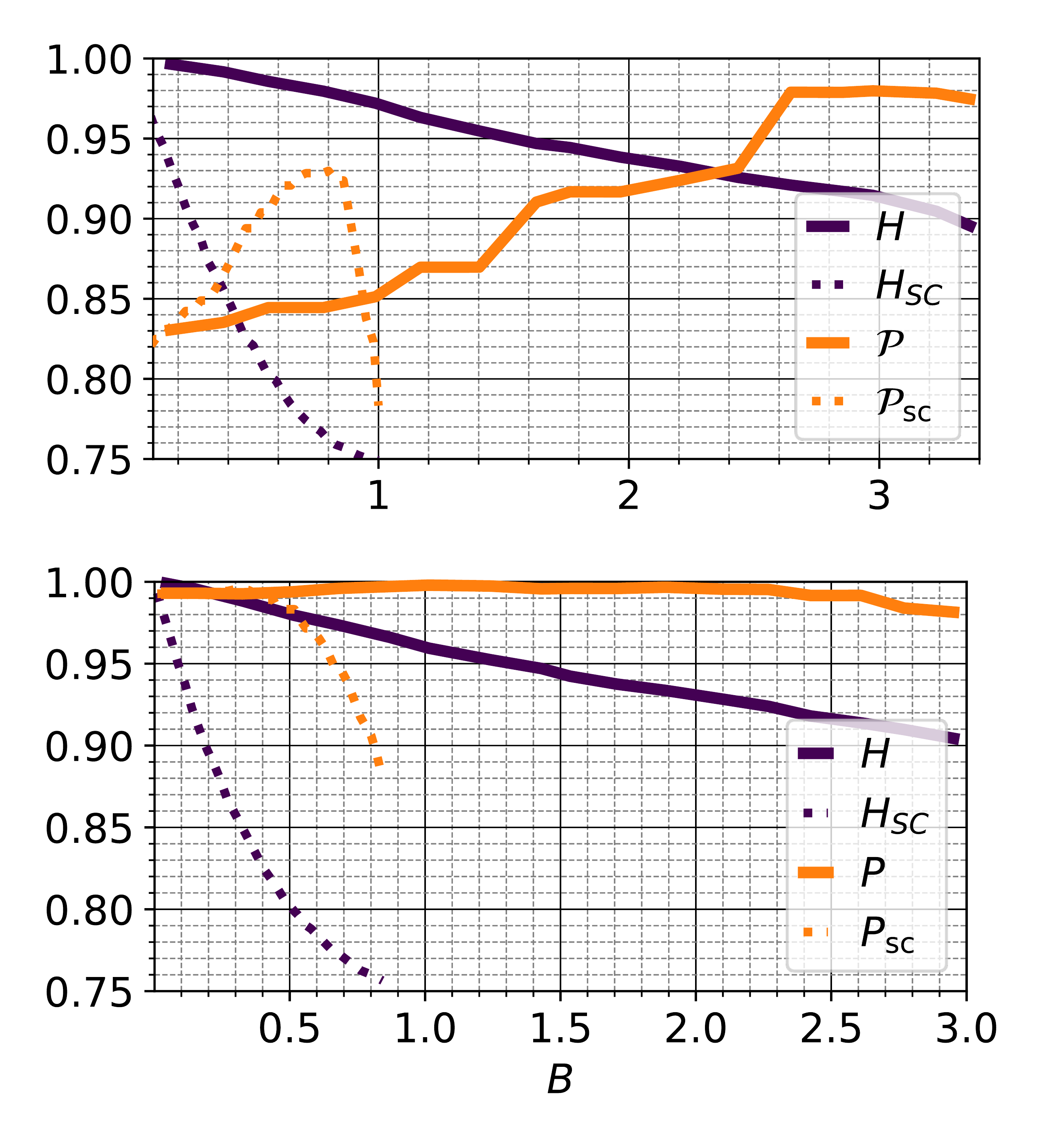}
	  \put(-3,76){\textbf{(a)}}	
    \put(-3,37){\textbf{(b)}}	

\end{overpic}
\caption{
(a) Maximum heralding efficiency as a function of relative brightness for the two-crystal (blue solid line) and single-crystal (blue dashed line) configurations, showing an improvement of more than one order of magnitude in the brightness--heralding-efficiency trade-off. The orange lines show the spectral purity of the corresponding optimized configurations for the two-crystal (solid line) and single-crystal (dashed line) architectures, indicating that high spectral purity can also be achieved in the tight-focusing regime. (b) Same analysis for the case of aperiodic poling with Gaussian phase matching, showing a similar improvement in the trade-off. In this case, the spectral purity remains nearly constant over the considered brightness range, in contrast to the single-crystal configuration.}
\label{fig:1D_B_H_P}

\end{figure}

We now quantify the trade-off between pair-collection probability and heralding efficiency for the two-crystal configuration and compare it with the single-crystal case. For each fixed value of $B$, we optimize $H$ over $\xi_\mathrm{p}$, $\xi_\mathrm{s}$, and $\phi_T$. The result is shown in Fig.\,\ref{fig:1D_B_H_P}(a), where $H$ is plotted as a function of $B$ (solid blue line). Heralding efficiencies above $0.9$ remain accessible up to $B<3.4$, indicating a strong relaxation of the usual trade-off. Moreover, $H>0.99$ is achieved at $B\approx0.4$, corresponding to nearly a 20-fold enhancement over the single-crystal case. The corresponding single-crystal curve is shown for comparison (dashed blue line).

We also compute the spectral purity for each optimized $(\xi_\mathrm{p},\xi_\mathrm{s},\phi_T)$ configuration (orange line). Spectral purities up to $\mathcal{P}\approx0.98$ are obtained in the tighter-focusing regime. Although the optimization here targets heralding efficiency, slight parameter retuning can increase the purity further, approaching unity without significantly reducing $H$ (as it was shown in Fig.\,\ref{fig:B_H_Spect} for the configuration represented by the magenta markers). 

\subsection{Trade-off comparison using aperiodically-poled crystals } \label{subsec:Tradeoff_AP}

An alternative route to high spectral purity is offered by a Gaussian nonlinear profile, $\chi^{(2)}(z)=\exp\left(-z^2/2\sigma^2\right)$, e.g using domain-engineered NLCs, typically referred as aperiodically-poled NLCs. SPDC with this Gaussian nonlinearity was recently shown to exhibit nearly the same brightness--heralding-efficiency trade-off as a constant nonlinearity, while the spectral purity decreases under tight focusing due to the increasing asymmetry of the spectral amplitude\,\cite{Sevilla2025:Efficient}. Here, we apply this phase-matching profile in the proposed cascaded configuration. We set $\sigma=L/4$ and compute the maximum $H$ as a function of $B$, together with the corresponding spectral purity. The results are shown in Fig.\,\ref{fig:1D_B_H_P}(b), where $H$ and $\mathcal{P}$ are plotted as solid blue and orange lines, respectively, while the corresponding single-crystal performance is shown with dashed lines.

We find that the brightness--heralding-efficiency trade-off is improved by more than one order of magnitude; that is, $B$ is increased more than ten times for equal $H$, while the spectral purity remains above $0.99$ over most of the parameter range. In contrast, for the single-crystal configuration, $\mathcal{P}$ drops rapidly for $B>0.4$. We note that the use of aperiodic poling in our cascaded architecture is beneficial particularly in high heralding-efficiency regime ($H>0.92$). However, for tighter focusing configuration, we find no significant benefit with respect to periodic poling.


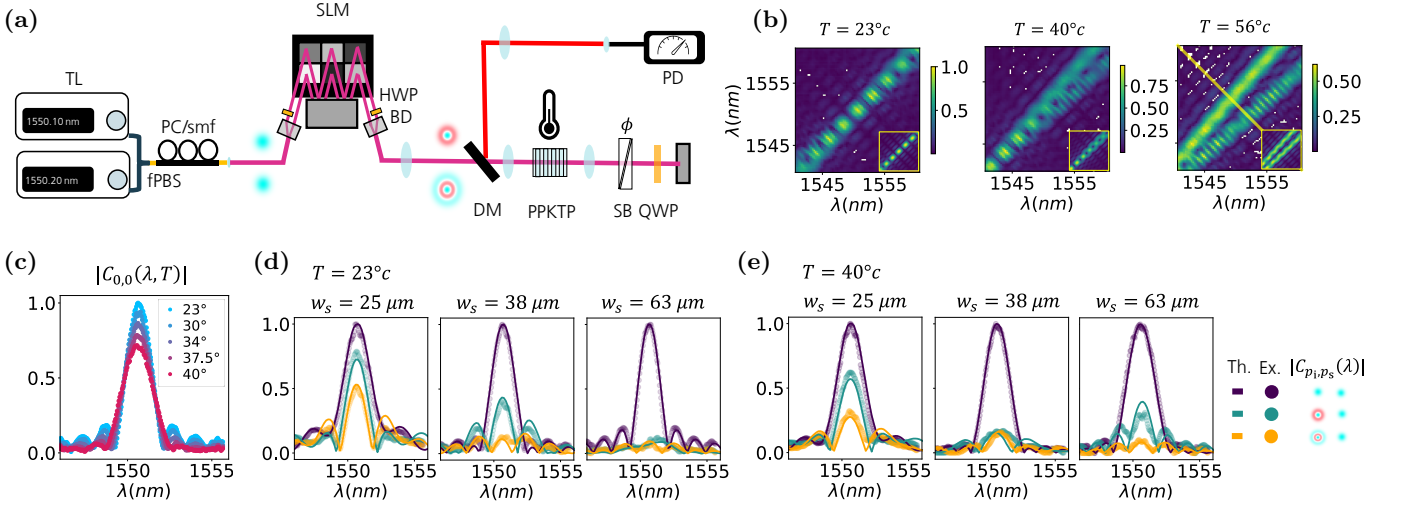
\begin{figure*}[t]
\centering
\begin{overpic}[unit=1mm,width=1\textwidth]{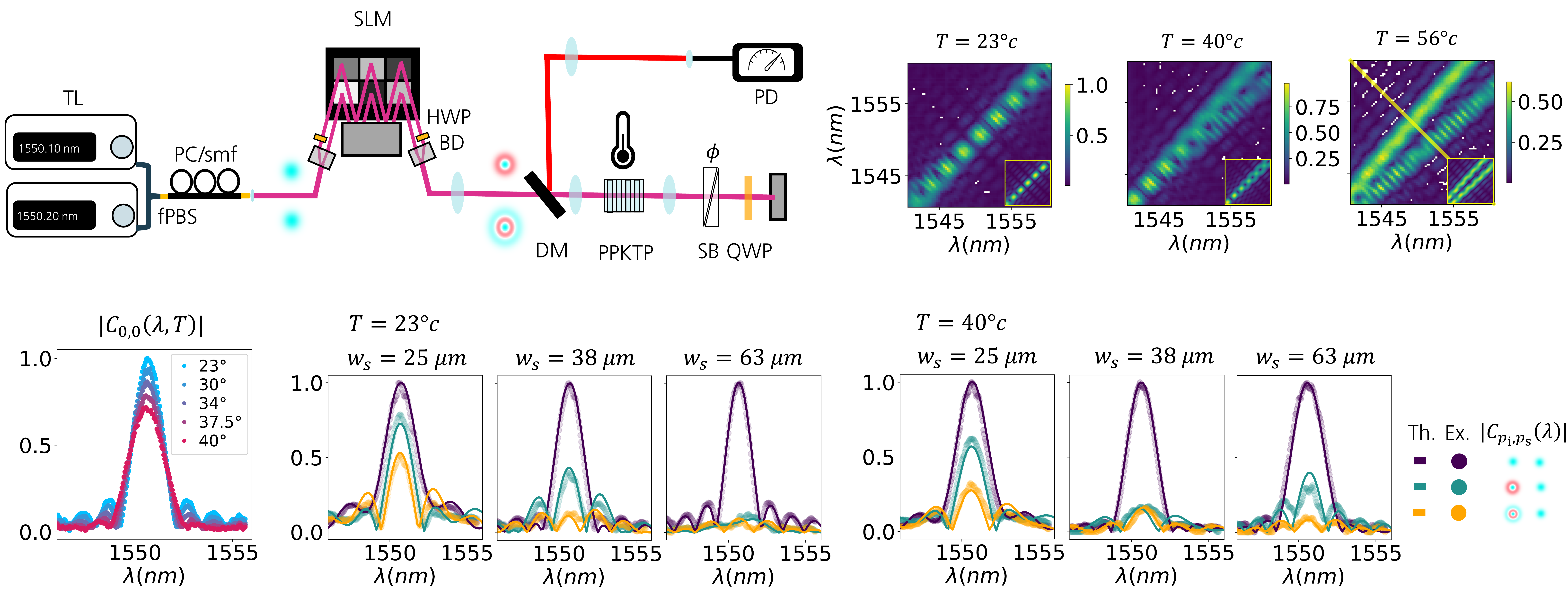}
	 \put(-1,65){\textbf{(a)}}	
	\put(98,65){\textbf{(b)}}
	   \put(-1,33){\textbf{(c)}}
       \put(32,33){\textbf{(d)}}
       \put(96,33){\textbf{(e)}}
\end{overpic}
\caption{(a) Experimental setup for the spatio-spectral characterization of the cascaded SPDC source using sum-frequency generation.
TL, tunable laser; PC, polarization controller; SMF, single-mode fiber; SLM, spatial light modulator; BD, beam displacer; HWP, half-wave plate; DM, dichroic mirror; SB, Soleil--Babinet compensator; QWP, quarter-wave plate; PD, photodiode. The signal and idler seed beams are independently prepared in wavelength and spatial mode and overlapped inside the nonlinear interferometer. The generated sum-frequency signal is detected at the output. (b) Experimental magnitude of the phase-matching function at three different temperature setpoints. The insets show the theoretical prediction taking into account the dispersion introduced by the lens inside the interferometer. (c) Experimental magnitude of the spectral amplitude for the Gaussian mode, $|C_{0,0}(\Omega)|$, recorded at temperatures between $23^\circ\mathrm{C}$ and $40^\circ\mathrm{C}$, showing a significant reduction of the sidelobes. (d),(e) Experimental magnitude of the spectral amplitudes for different mode combinations, $|C_{p_\mathrm{s},p_\mathrm{i}}(\Omega)|$, at $23^\circ\mathrm{C}$ and $40^\circ\mathrm{C}$, respectively. At each temperature, different focusing configurations for the signal and idler photons are used.}
\label{fig:setup_experiment}
\end{figure*}

\section{Experiment}\label{sec:experiment}

The experimental setup used to validate our model is shown in Fig.\,\ref{fig:setup_experiment}(a). We employ SFG measurements, in which the spatio-spectral state of the signal and idler fields is prepared and the upconverted field is measured in the Gaussian mode via single-mode-fiber coupling. This allows us to directly probe $|\Phi_{p_\mathrm{s},p_\mathrm{s}}(\Omega_s,\Omega_i)|$ up to a normalization factor. The signal and idler fields are generated by two fiber-coupled tunable lasers (TL, Santec TSL-770). Each laser is connected to a different output port of a fiber polarizing beam splitter (fPBS), allowing independent wavelength tuning of the two polarizations. The co-propagating output beams are then collimated, separated by a beam displacer (BD), and sent to a spatial light modulator (SLM) for beam shaping. Because the SLM is polarization sensitive, the polarization of one beam is rotated with a half-wave plate (HWP). In this way, the spatial modes of the signal and idler can be prepared independently using multi-plane light conversion (MPLC)\,\cite{Hiekkamaki:2019_Near-perfect}.

The signal and idler fields are subsequently recombined using a second HWP--BD combination and imaged into the center of the nonlinear crystal (NLC) with an $8\times$ demagnification. The NLC is a PPKTP crystal of length $L=20\,\mathrm{mm}$ poled for type-II interaction at $\lambda_\mathrm{s,i}=1550\,\mathrm{nm}$. The signal, idler, and SFG fields then co-propagate and are imaged back into the center of the crystal using a lens--mirror configuration. Double pass through a quarter-wave plate (QWP) oriented at $45^\circ$ rotates the polarization of the signal and idler fields by $90^\circ$. Note that the polarization of the co-propagating pump is unchanged since the QWP acts as a HWP at the pump wavelength. A custom-made Soleil--Babinet compensator (SB) adjusts the relative phase between the three fields, $\phi=\phi_\mathrm{p}-\phi_\mathrm{s}-\phi_\mathrm{i}$. After the second pass through the crystal, the SFG field is separated by a dichroic mirror (DM), coupled into a single-mode fiber, and detected with a power meter. In the reported measurements, the SFG waist inside the crystal is set to $w_\mathrm{p}\approx95\,\mu\mathrm{m}$, corresponding to a focusing parameter of $\xi_\mathrm{p}\approx0.156$.

First, we measure the magnitude of the phase-matching function, $|\Phi_{0,0}(\Omega_\mathrm{s},\Omega_\mathrm{i})|$, for Gaussian signal and idler modes ($p_\mathrm{s}=p_\mathrm{i}=0$) with $w_\mathrm{s}\approx63\,\upmu\mathrm{m}$ ($\xi_\mathrm{s}\approx0.72$), for three different crystal temperatures. The results are shown in Fig.\,\ref{fig:setup_experiment}(b). As the temperature increases, the PMFs associated with the two crystals shift progressively away from one another. At $T=23^\circ\mathrm{C}$, they overlap almost completely at $1550.64\,\mathrm{nm}$, whereas at $T=56^\circ\mathrm{C}$ they are largely separated; this latter case is used to estimate the loss inside the interferometer.

In addition, a periodic modulation along the $(\Omega_\mathrm{s}+\Omega_\mathrm{i})$ axis is observed. It is most pronounced at $T=23^\circ\mathrm{C}$, where the overlap between the two PMFs is maximal. The slower modulation originates from the dispersion introduced by the lens and gradually disappears as the two PMFs separate ($T=56^\circ\mathrm{C}$). The corresponding phase term is given by $H(\Omega_\mathrm{s},\Omega_\mathrm{i})=(K_\mathrm{p}-K_\mathrm{s}-K_\mathrm{i})d$, as introduced in Eq.\,\ref{eq:NLI_2}, where $K_j$ and $d$ denote the wave vectors in the lens material and the lens thickness, respectively. For an isotropic material, the group velocities of signal and idler are equal, so that $H(\Omega_\mathrm{s},\Omega_\mathrm{i})=[K_0+(u_\mathrm{p}^{-1}-u_\mathrm{s}^{-1})(\Omega_\mathrm{s}+\Omega_\mathrm{i})]d$. The insets show the corresponding theoretical PMFs calculated from the lens specifications provided by the manufacturer. A second, faster modulation arises from Fabry--P\'erot interference in the QWP, which is antireflection-coated only for the telecom wavelength. As a result, it affects only the SFG wavelength. These modulations could be substantially reduced by using appropriate antireflection coatings and thinner, less dispersive lenses, or by replacing the lens with a concave mirror. However, to probe $|\Phi_{p_\mathrm{s},p_\mathrm{i}}(\Omega)|$, it is sufficient to consider only the $(\Omega_\mathrm{s}-\Omega_\mathrm{i})$ axis (yellow cross-section in Fig.\,\ref{fig:setup_experiment}(b)), where these effects do not contribute.

We now turn to the experimental validation of our results, starting with the spectral purity optimization via destructive interference of the sidelobes of the sinc function. Figure\,\ref{fig:setup_experiment}(c) shows the measured magnitude of the spectral amplitude for the Gaussian mode, $|C_{0,0}(\Omega)|$, recorded at temperatures between $23^\circ\mathrm{C}$ and $40^\circ\mathrm{C}$. For this measurement, $\phi$ was adjusted to compensate for the $\phi_T$ introduced when changing the temperature of the crystal. As expected, increasing the temperature leads to a pronounced suppression of the sidelobes, accompanied by a broader bandwidth and a reduced enhancement factor.

Finally, we measure the magnitudes of the spectral amplitudes for different mode combinations, $|C_{p_\mathrm{s},p_\mathrm{i}}(\Omega)|$, at two temperatures. We compare several focusing conditions for the SPDC photons to assess how accurately the model captures the contribution of each joint spatial mode. Figure\,\ref{fig:setup_experiment}(d) shows the results at $T=23^\circ\mathrm{C}$ for beam waists of $w_\mathrm{s}=25\,\upmu\mathrm{m}$, $38\,\upmu\mathrm{m}$, and $63\,\upmu\mathrm{m}$. The three panels reveal how strongly both the relative weight and the spectral shape of the joint spatial modes depend on the focusing conditions. The dotted curves correspond to the experimental data and the solid curves to the theoretical predictions, showing very good agreement. The third case corresponds to the regime discussed above, in which the higher-order mode combinations are strongly suppressed. Figure\,\ref{fig:setup_experiment}(e) shows the corresponding results at $T=40^\circ\mathrm{C}$, where similarly good agreement with the model is observed. Notably, a more optimal heralding efficiency configuration shifted from $w_\mathrm{s}=63\,\upmu\mathrm{m}$ to $w_\mathrm{s}=38\,\upmu\mathrm{m}$.

Overall, the experimental results are in good agreement with the theoretical model. They confirm the temperature-dependent reshaping of the spectral amplitudes, as well as the mode-selective enhancement and suppression induced by nonlinear interference. These observations support our interpretation of the improved brightness, heralding efficiency, and spectral purity in terms of interference between different spatio-spectral modes.

\section{Discussion}\label{sec:discussion}

In summary, we have shown that nonlinear interference improves the brightness-heralding trade-off of type-II quasi-phase-matched SPDC sources by more than one order of magnitude. Equivalently, for a fixed heralding efficiency, the brightness is enhanced by more than a factor of ten. In addition, we have identified strategies for improving the spectral purity by exploiting nonlinear interference in a two-crystal configuration with intermediate polarization rotation, phase control, and temperature tuning.

Among the most attractive operating regimes, we identify configurations that combine excellent source performance in different ways. In particular, we find regimes with extremely high heralding efficiency ($H>0.99$) together with a 20-fold brightness enhancement relative to a single crystal. We also find operating points with spectral purities of $0.98$--$0.99$ and relative brightness exceeding $3$, corresponding to an improvement of about 13-fold over the single-crystal case while still maintaining heralding efficiencies above $0.9$. In addition, we propose a simple route to simultaneously high spectral purity and high heralding efficiency by tuning the crystal temperature, while still retaining a twofold brightness enhancement. 

For Gaussian phase matching, we find that the usual trade-off between spectral purity and brightness is essentially removed. This makes the source especially attractive in the tight-focusing regime, in contrast to the single-crystal case, where the spectral purity is strongly degraded. However, in this regime aperiodically poled crystals do not offer a significant additional advantage over periodically poled crystals, since both can achieve similarly high spectral purity.

These results enhance the practical potential of bulk SPDC sources by relaxing the usual trade-offs between brightness, heralding efficiency, and spectral purity. More broadly, they show that nonlinear interference provides a powerful tool for filter-free, mode-selective engineering at the source level, requiring only relatively simple modifications to already available nonlinear-crystal sources. This could allow existing quantum-optics experiments to substantially improve their performance without adopting more complex source architectures. Our work is particularly relevant for applications such as loophole-free Bell tests and device-independent quantum key distribution, where high heralding efficiency must be combined with large pair rates. It is also relevant for experiments based on high-purity heralded photons and multi-photon interference, including Gaussian boson sampling, entanglement swapping, and related photonic quantum-information protocols.


Beyond source optimization, the formalism could also be used to study nonlinear interferometers in bulk systems while accounting for the full spatial--spectral degree of freedom, providing a route for engineering more complex spatial--spectral biphoton states. In this sense, nonlinear interference complements periodic and aperiodic poling as a simple strategy for tailoring the quantum states generated in nonlinear media.



\section{Funding}
The authors acknowledge support by the Carl-Zeiss-Stiftung within the Carl-Zeiss-Stiftung Center for Quantum Photonics (CZS QPhoton) under the project ID P2021-00-019. This work has received funding from the German Federal Ministry of Research, Technology and Space (BMFTR) within the PhoQuant project. Funded by the Deutsche Forschungsgemeinschaft (DFG, German Research Foundation) – Project-ID 398816777 – SFB 1375.

\section{acknowledgements}
 C.S.G and F.S. are members of the Max Planck School of Photonics, supported by the German Federal Ministry of Education and Research, the Max Planck Society, and the Fraunhofer Society. 

\section{Disclosures}
The authors declare no conflicts of interest

\section{Data Availability}
The data underlying the results presented in this paper can be obtained from the authors upon reasonable request.

\appendix

\section{Theory}
\subsection{Laguerre-Gauss modes}
The definition for the Laguerre-Gauss (LG) modes used is given by 
\begin{align}\label{eq_sup:LG_function}
     \mathrm{LG}_{p}^{\ell}(\rho,\varphi)=\mathrm{exp}\big(\frac{-\rho^2w^2}{4} +i\ell\varphi\big)\sum_{u=0}^pT_u^{p,\ell}\rho^{2u+|\ell|},
\end{align}
with
\begin{equation}
         T_u^{p,\ell} = \sqrt{\frac{p!\,(p+|\ell|)!}{\pi}}\,
   \biggr(\frac{ w}{\sqrt{2}}\biggl)^{2u+|\ell|+1}\,\frac{(-1)^{p+u}(i)^{\ell}}{(p-u)!\,(|\ell|+u)!\,u!}
\end{equation}

\subsection{Propagation of SPDC through an optical system}
\begin{figure}[b]
    \centering

    \begin{overpic}[unit=1mm,width=0.5\textwidth]{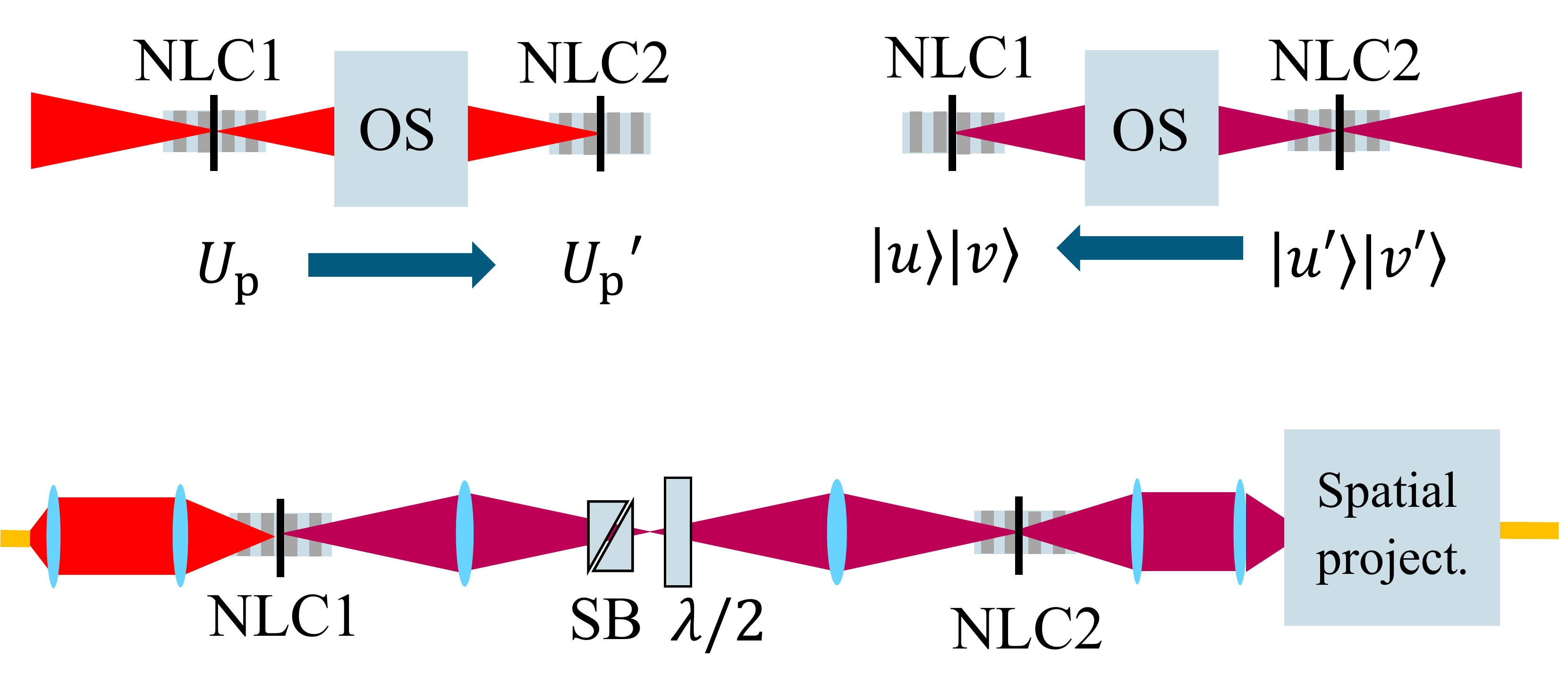}
	  \put(-2,38){\textbf{(a)}}	
    \put(43,38){\textbf{(b)}}	
    \put(-2,16){\textbf{(c)}}	

\end{overpic}
\caption{(a) Illustration of the pump propagation. (b) Illustration of the back propagation of the signal and idler fields. (c) Optical system employed in the experiment.}
    \label{fig_app:ABCD}
\end{figure}

Here, we describe the methodology for simulating our experimental setup. In general, it can be used to simulate any optical system that can be described with the ABCD formalism. 

First, we define the spatial structure of the pump mode $U_\mathrm{p}$ in the center of the first nonlinear crystal (NLC1). The pump mode travels through an optical system (OS) and reaches the center of the second nonlinear crystal (NLC2) as depicted in Fig.\,\ref{fig_app:ABCD}(a). The mode is transformed by the OS to $U_\mathrm{p}'$. 

Second, we define the spatial modes for signal and idler $\ket{u'}$ and $\ket{v'}$, respectively, in the center of NLC2. When back-propagating through the optical system to the center of NLC1, the modes transform to $\ket{u}$ and $\ket{v}$, respectively. This is depicted in Fig.\,\ref{fig_app:ABCD}(b). 

These modes can be described as a superposition of LG modes as $U_\mathrm{p}=\sum_{\ell_\mathrm{p},p_\mathrm{p}}A_{\ell_\mathrm{p},p_\mathrm{p}}LG_{\ell_\mathrm{p},p_\mathrm{p}}$, $\ket{u} =\sum_{\ell_\mathrm{s},p_\mathrm{s}}A'_{\ell_\mathrm{s},p_\mathrm{s}}\ket{\ell_\mathrm{s},p_\mathrm{s}}$ and $\ket{v} =\sum_{\ell_\mathrm{i},p_\mathrm{i}}A''_{\ell_\mathrm{i},p_\mathrm{i}}\ket{\ell_\mathrm{i},p_\mathrm{i}}$. 

The phase matching function (PMF) of that configuration can  be calculated through:

\begin{equation}
    C_{u,v}(\Omega_\mathrm{s}, \Omega_\mathrm{i}) = \sum_{\ell_\mathrm{p},p_\mathrm{p}}\sum_{\ell_\mathrm{s},p_\mathrm{s}}\sum_{\ell_\mathrm{i},p_\mathrm{i}}\mu_{\ell_\mathrm{p},p_\mathrm{p},\ell_\mathrm{s},p_\mathrm{s},\ell_\mathrm{i},p_\mathrm{i}}C^{\ell_\mathrm{p},\ell_\mathrm{s},\ell_\mathrm{i}}_{p_\mathrm{p},p_\mathrm{s},p_\mathrm{i}}(\Omega_\mathrm{s}, \Omega_\mathrm{i})\,, 
\end{equation}

where $\mu_{\ell_\mathrm{p},p_\mathrm{p},\ell_\mathrm{s},p_\mathrm{s},\ell_\mathrm{i},p_\mathrm{i}}=A_{\ell_\mathrm{p},p_\mathrm{p}}A'_{\ell_\mathrm{s},p_\mathrm{s}}A''_{\ell_\mathrm{i},p_\mathrm{i}}$. Similarly,  $C_{u,v}(\Omega_\mathrm{s}, \Omega_\mathrm{i})$ is calculated. The result of the nonlinear interference is:

\begin{align}\label{eq:NLI_ABCD}
    \Phi_{u,v}(\Omega_\mathrm{s},\Omega_\mathrm{i})=[C_{u,v}(\Omega_\mathrm{s},\Omega_\mathrm{i})e^{-i\frac{\Delta k(\Omega_\mathrm{s},\Omega_\mathrm{i}) L}{2}}]_{f}e^{i(-\frac{\Delta k L}{2}+\phi)}\nonumber&\\
    +C_{u', v'}(\Omega_\mathrm{s},\Omega_\mathrm{i}).
\end{align}

Now we calculate the effect of OS on $U_\mathrm{p}$, $\ket{u}$ and $\ket{v}$. Since these fields are superposition of LG modes, we can use the ABCD Law for LG beams\,\cite{Tache:87}. Note that the LG beams are a function of the waist parameter $w$, but in general, it is a function of the beam parameter $q$. Using the ABCD matrix formalism, we can easily calculate the transformation of $q\rightarrow{q'}$. Note, this transformation is mode-order dependent through the accumulated Gouy phase of OS. However, this term is ideally zero for a 4-f system.    
LG bases defined with different beam parameter, are no longer orthogonal in the radial index and one can be written as a coherent superposition of the other. That is $LG_{\ell,p}(q)=\sum_{p'}B_{p'}LG_{l,p'}(q')$. This was studied by Vallone\,\cite{Vallone:17} finding an analytical expression to map different LG families with different $q$-parameter. This reduces the simulation of a more realistic OS, such as thick lenses, focusing shifts between pump and spdc wavelengths, etc., to a few matrix multiplications. 

As the basis for the simulation of our setup in Fig.\,\ref{fig_app:ABCD}(c), we consider $C^{\ell_\mathrm{p},\ell_\mathrm{s},\ell_\mathrm{i}}_{p_\mathrm{p},p_\mathrm{s},p_\mathrm{i}}(\Omega_\mathrm{s}, \Omega_\mathrm{i})$ for the case of type II SPDC in a PPKTP of length $L=20\,\mathrm{mm}$. The pump and SPDC modes were defined with respect to the waist parameters $w_\mathrm{p}=90\,\mu\mathrm{m}$ and $w_\mathrm{s}=50\,\mu\mathrm{m}$. We run the radial index for the SPDC photons from $0$ to $8$, to cover all the investigated collection waists. As the pump was fixed, we only consider $p_\mathrm{p}=0$ and $1$.

\section{Numerical simulation }

The range and step size of the parameters used in the calculations are given in the following table:

\begin{table}[h]
    \centering
    \begin{tabular}{|c|c|c|}
        \hline
        Parameter & Range & Step \\
        \hline
        $log_{10}(\xi)$ & -2 to 1 & 0.05 \\
        \hline
        $D\Omega L$ & -406.12 to 59.14 & 1.6 \\
        \hline
    \end{tabular}
    \label{tab:sampling_T2}
\end{table}

\nocite{*}
\bibliography{Biblio}

\end{document}